\documentclass[12pt]{article}

\usepackage{latexsym,cmmib57}
\usepackage{amsmath}
\usepackage{amsthm}
\usepackage[psamsfonts]{amssymb,amsfonts,eucal}
\usepackage{graphicx}
\usepackage{url}
\usepackage{times}
\usepackage{setspace}
\newcommand{\bbrd}[1]{\mbox{\rm{I}\kern-.1667em{#1}}}

\newsavebox{\fmbox}

\newcounter{algocnt}

\newcommand{\wlen}{\ell}

\begin{document}
\title{Reconsidering the significance of genomic word frequency}
\author{Mikl\'os Cs\H{u}r\"os\thanks{%
		Department of Computer Science and Operations Research,
		Universit\'e de Montr\'eal, 
		CP 6128, succ.\ Centre-Ville, Montr\'eal, Qu\'ebec H3C 3J7, Canada.}
	\and Laurent No\'e\thanks{%
		Laboratoire d'Informatique Fondamentale de Lille,
		B\^at.\ M3, 
		59655 Villeneuve d'Ascq C\'edex, France.}
	\and Gregory Kucherov$^{\thefootnote}$}

\maketitle

Determining
what constitutes unusually frequent and rare
in a genome is a fundamental and ongoing 
issue in genomics~\cite{Karlin.symposium}.
Sequence motifs may be 
frequent because 
they appear in mobile, structural or regulatory elements.
It has been suggested 
that some recurrent sequence 
motifs indicate hitherto unknown or poorly understood 
biological phenomena~\cite{Pyknon}.
We propose that the distribution of
DNA words in genomic sequences 
can be primarily characterized by a 
double Pareto-lognormal distribution,
which explains lognormal and power-law 
features found across all known genomes.
Such a distribution may be the result of 
completely random sequence evolution by 
duplication processes.
The parametrization of 
genomic word frequencies allows for an 
assessment of significance for frequent or rare 
sequence motifs.

The simplest type of sequence motif 
is a DNA word of a fixed length~$\wlen$, called 
an {\em $\wlen$-mer}. 
The number of occurrences of a word~$w$ 
is denoted by~$N(w)$, and the 
distribution of~$N(w)$
across all $\wlen$-mers is called the {\em word frequency distribution}
or {\em spectrum}.
The statistical significance 
of a word's unusual abundance (or rarity)
is assessed by referring to a null model 
of random sequences.
Standard null models 
include sequences of independent and 
identically distributed letters 
({\em Bernoulli model}\/)
and low-order 
Markov models~\cite{ReinertSchbathWaterman}.
In such random text models, 
over- and underrepresentation 
of short words (typically, $8\le\wlen\le 16$) 
are evaluated by using Poisson or Gaussian
approximations, implying a 
rapidly decreasing tail in 
the spectrum.
It is also customary to use localized random shuffling 
of the studied genome sequence
in order to preserve large-scale 
compositional heterogeneity.
Empirical word frequency distributions
in shuffled sequences 
also have a light tail.

\begin{figure}
\centerline{\includegraphics[height=0.4\textheight]{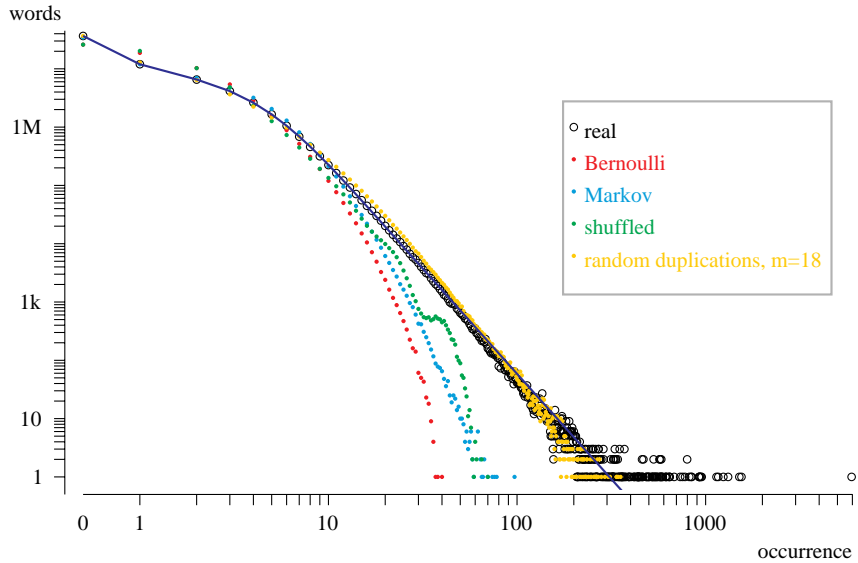}}
\caption{13-mer frequencies in repeat-masked human chromosome 5, and in 
	random sequences of same length.}\label{fig:random}
\end{figure}

\begin{figure}
\centerline{\includegraphics[height=0.4\textheight]{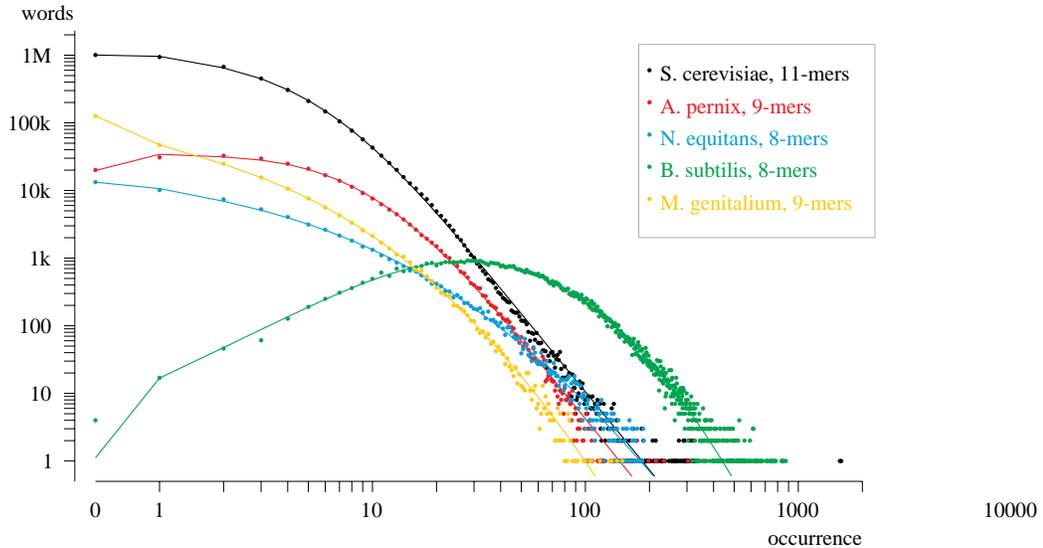}}
\caption{Spectra in smaller genomes. For each full genome (concatenated chromosome sequences
	if necessary), the $\ell$-mer frequency distribution is shown by dots, and the fitted 
	DPL distribution by a solid line. Notice the power law in the lower tail of the 
	B.~subtilis genome spectrum.}\label{fig:prok}
\end{figure}

\begin{figure}
\centerline{\includegraphics[height=0.4\textheight]{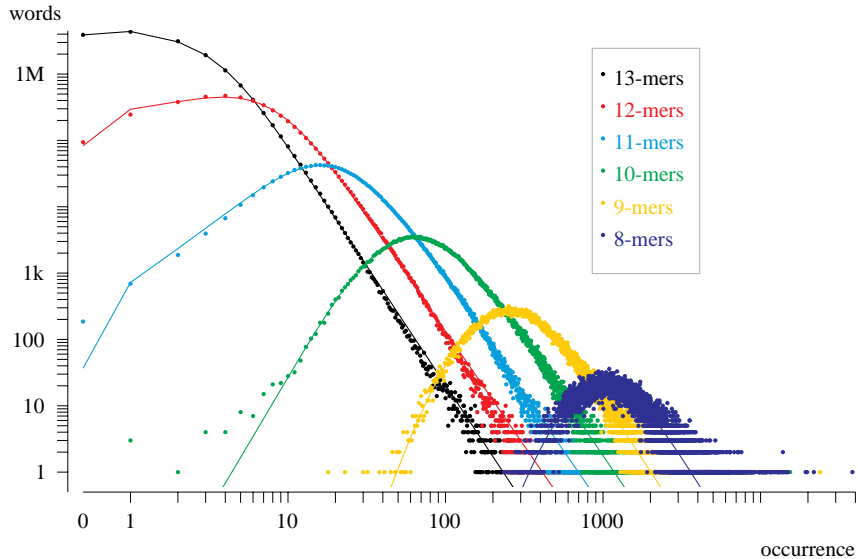}}
\caption{Spectra of CpG-free $\ell$-mers on repeat-masked human chromosome 12.}\label{fig:hs12}
\end{figure}

In reality, genomic word frequency distributions
have a prominent heavy tail,
which is not captured by random text models (Fig.~\ref{fig:random}).
Depending on the genome length~$L$
and its relative size with respect to
the number of~$\wlen$-mers~$4^{\wlen}$, 
the spectrum may show a power-law 
decrease on the left or right,
or have a lognormal shape.
The heavy tail on the right cannot be entirely
attributed to mobile elements, as 
it is present even in 
repeat-masked vertebrate 
and many smaller genomes 
(Figures~\ref{fig:prok} and~\ref{fig:hs12}). 
Consequently, 
random text models and shuffling 
tend to underestimate the probability 
of frequent words in 
long sequences.

To this day, there has been no exact characterization 
of genomic spectra,
aside from the observation of power-law behavior for 
certain word sizes~\cite{Luscombe.power,Mantegna,MartindaleKonopka}
in the right-hand tail.
This note aims to point out that 
a parametric distribution describes 
word frequencies extremely well 
in prokaryotic and eukaryotic genome sequences. The distribution in question is 
the so-called {\em double Pareto-lognormal} (DPL) distribution~\cite{ReedJorgensen}.
It fits many real-life size distributions,
including that of wealth in society, human settlement sizes, and file sizes on the Internet.
The distribution has four parameters: $\alpha>0$,
$\beta>0$, $\nu$ and $\tau>0$;
it has a power-law (Pareto) tail to the left and to the right, with slopes characterized 
by the parameters~$\beta$ and~$\alpha$, respectively; in the middle, its shape is dominated by a lognormal distribution
with parameters~$\nu$ and~$\tau$. 

Figure~\ref{fig:prok} illustrates that a single DPL distribution's four parameters 
can be adjusted to describe spectra 
across hundreds or thousands of word frequencies in non-vertebrate genomes. 
(More examples for different spectra are shown on
the web site \url{http://www.iro.umontreal.ca/~csuros/spectrum/}.)
For very short words with respect to the genome length~$L$
(about $8\le \wlen\le (\log_4 L)-4$), 
the spectrum has mainly a lognormal shape. As~$\wlen$
increases, the mode of the lognormal component shifts downwards and 
the lower power-law tail becomes more and more discernible, followed by the appearance
of the upper power-law tail. 
For long words (from around $\wlen>(\log_4 L)+2$), 
the upper power-law tail dominates the spectrum. 
Given that compact genomes are mostly composed of coding sequences, 
the upper power-law tail is 
another manifestation of the
power law for protein domain occurrences~\cite{Luscombe.power}.

\begin{figure}
\centerline{\includegraphics[height=0.4\textheight]{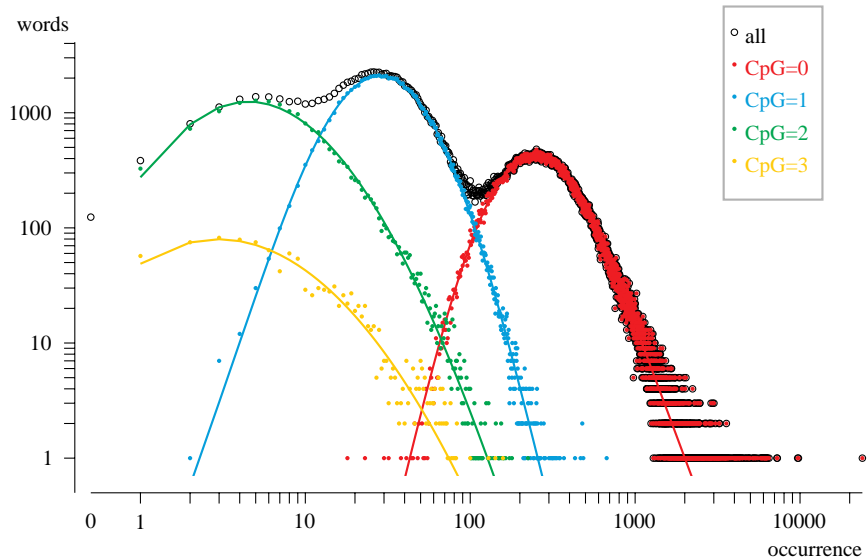}}
\caption{9-mer frequencies in non-repeat annotated regions of human chromosome 12.
	The spectrum can be decomposed by CpG content into a few DPL distributions, 
	shown by solid lines.}\label{fig:CpG}
\end{figure}

\begin{figure}
\centerline{\includegraphics[height=0.4\textheight]{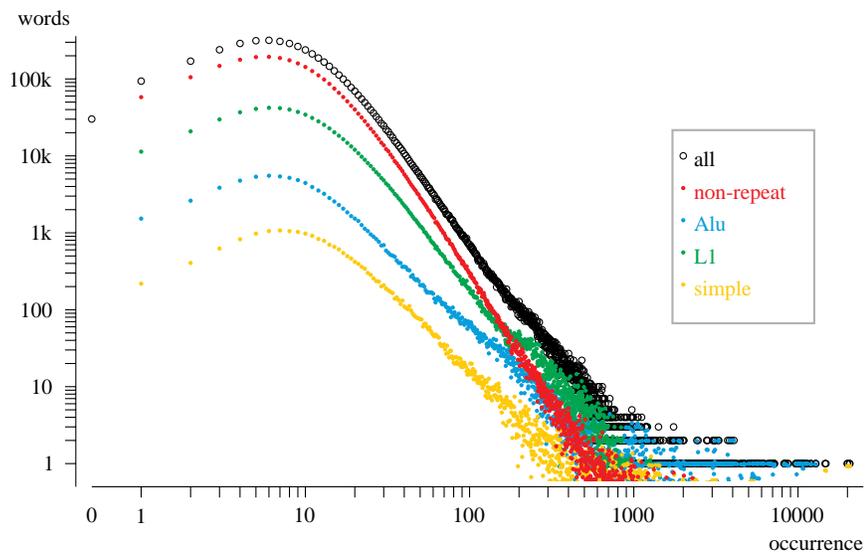}}
\caption{Decomposing the contribution of different repeat families 
	in the spectrum of CpG-free 12-mers along human chromosome 12.
	Most of the deviation from DPL in the right-hand tail is caused by Alu and L1 families and simple repeats. 
	}\label{fig:repeats}
\end{figure}

In organisms with strong dinucleotide bias, such as for CpG in 
vertebrates or for ApT in honey bee~\cite{genome.bee}, 
the spectrum can be decomposed into multiple DPL distributions
by dinucleotide content (Figure~\ref{fig:CpG}). 
Spectra for words without CpG dinucleotides 
in repeat-masked vertebrate sequences have a 
marked DPL shape (Fig.~\ref{fig:hs12}).
We illustrate 
the analysis of a large genomic spectrum with the example of 
human chromosome~12, 
which is typical of the human genome with respect to repeat element
distribution and cytosine-guanine content~\cite{hs12.finish}. 
Abundant repeat elements may cause deviations from the DPL distribution,
which may be the basis of their identification 
using a DPL null model, 
but often they are absorbed in the fundamental DPL curve (Figure~\ref{fig:repeats}). 
Table~\ref{tbl:chr12.tail} 
analyzes the composition of the spectrum's tail in 
human chromosome~12. 
The contribution of non-repeat sequences 
in the tail decreases when moving toward higher word frequencies, 
but it levels off at about 25\%. 

\begin{table}
\small
\begin{tabular}{rlllllll}
$n$ & 12-mers (a) & sequence (b) & non-repeat (c) &  SINE (d)& LINE (e)& LTR (f) & other (g)\\
\hline
12		&19.61	&69.6	&47.2	&16.7	&21.4	&8.4	&6.3\\
25		&5.34	&39.7	&41.5	&21.3	&22.6	&7.6	&7.0\\
30		&3.67	&33.9	&39.6	&23.1	&22.9	&7.3	&7.2\\
50		&1.30	&22.6	&34.0	&28.8	&23.2	&6.5	&7.5\\
60		&0.91	&19.9	&32.1	&31.0	&23.1	&6.2	&7.5\\
100		&0.37	&14.7	&27.9	&36.7	&22.6	&5.4	&7.4\\
120		&0.28	&13.4	&26.8	&38.5	&22.3	&5.1	&7.3\\
200		&0.14	&10.6	&24.6	&43.0	&20.7	&4.7	&7.0\\
240		&0.10	&9.7	&24.2	&44.9	&19.4	&4.6	&6.9\\
\hline
\end{tabular}
\caption{Composition of the 12-mer spectrum's tail
in human chromosome 12.
Column (a) gives the 
fraction of words that occur at least~$n$ times;
column (b) lists the fraction of the genome sequence 
covered by such words. 
Columns (c--g) list the 
fraction of those word occurrences within
non-repeat regions, 
short interspersed elements, 
long interspersed elements, 
long terminal repeats, and 
other repeat elements (including DNA transposons, simple repeats,
low-complexity and 
tandem repeats), respectively. Fractions are expressed as percentages.}\label{tbl:chr12.tail} 
\end{table}

The power-law tail of gene family size distributions~\cite{Luscombe.power} 
can be explained by birth and death processes~\cite{Karev.BDIM.2002,Reed2004}. 
Similar arguments apply to genomic spectra.
Consider the occurrences of a particular word along the genome as a ``population.''
The population size is affected by mutational events, including
duplications, insertions, deletions and point mutations.
The population can increase by any copying mechanism, including 
segmental duplication and
retrotranscription. The population decreases if an occurrence 
is destroyed by some mutation.
Point mutations can create new populations, but so 
can insertion (at the insertion boundaries) and deletion (by bringing 
two halves of a word together). 
It is thus conceivable that a birth and death model appropriately 
models the spectrum's evolution.
In order to illustrate this point, we carried out 
a simulation 
experiment in which a DNA sequence evolved solely by copying. 
We iteratively expanded an initial random DNA sequence
by selecting 
a contiguous piece of a fixed length~$m$ in each iteration,
and copying it back into the sequence at a random position. 
The resulting sequence exhibits the same heavy tail as the real-life sequences 
do (Figure~\ref{fig:random}). 
The parameters of the fitted DPL distribution are affected 
by the copy size~$m$, and further vary with the introduction of a point mutation 
process.
It is thus likely that the DPL distribution
of genomic spectra is yet another sign of 
ceaseless ``tinkering''~\cite{tinkering}
within the genome.

Heavy-tail distributions are signs of self-similarity, or long-range autocorrelation, 
which has been observed before in DNA sequences~\cite{LiKaneko,long.range.DNA},
and studied thoroughly in telecommunications engineering~\cite{multimedia.self-similar}. 
Long-range autocorrelation at the single nucleotide level can result from 
so-called expansion-randomization processes~\cite{universal.expansion.randomization},
which model sequence evolution by deletions, mutations and duplications. 

The birth and death model implies that some words 
occur often simply by chance, and not because of their functionality.
Words that are abundant at an early point of evolution tend to preserve 
their relative abundance in the course of random copy events,
in accordance with the principle of ``rich get richer'' 
underlying power-law distributions.
Therefore, even 
the high frequency of a particular word across many related 
species does not imply functionality on its own, 
as the word may have been 
frequent by chance in a common ancestor already. 
Association with genes is not necessarily a 
sign of functionality either,
since transcribed regions that
harbor a frequent word 
can help its propagation throughout 
the genome by retrotranscription. 

Our investigations show that the distribution of word frequencies can be 
well approximated by a parametric model, the double Pareto-lognormal distribution. 
Such a distribution may result from a long history of evolutionary tinkering:
copying, rearranging, deleting, and changing different parts of the genome.
The heavy tail of word frequency distributions means that findings of 
frequent motifs need to be assessed with extreme care, especially 
if their overrepresentation is related to word occurrences in random texts.

%[``unique'' ???]

{\small
\section*{Methods}
Words were counted only on one strand of the DNA sequences (the `plus' strand of the 
sequence file), with the exception of the 16-mers in the human genome, where both 
strands were scanned. We counted the occurrence of a word~$w$ 
if it appeared in a given sequence at some position $i..i+\wlen-1$,
without ambiguous nucleotides. The DPL distribution was fitted 
using its cumulative distribution function (cdf), which is
\begin{multline*} %begin{equation}\label{eq:dPln}
F(x) 
	= %\begin{aligned}[t] 
		%& 
		\Phi\Bigl(\frac{\ln x-\nu}{\tau}\Bigr) 
		+ %& 
		\frac{\alpha}{\alpha+\beta}
			x^\beta 
			e^{-\beta\nu+\beta^2\tau^2/2}
			\Phi\Bigl(-\frac{\ln x-\nu+\beta\tau^2}{\tau}\Bigr) \\*
		-  \frac{\beta}{\alpha+\beta}
			x^{-\alpha}
			e^{\alpha\nu+\alpha^2\tau^2/2}
			\Phi\Bigl(\frac{\ln x-\nu-\alpha\tau^2}{\tau}\Bigr)
	%\end{aligned}
\end{multline*} %end{equation}
for $x>0$ and $F(x)=0$ for $x\le 0$,
where $\Phi(\cdot)$ denotes the cdf of the standard normal distribution.
The spectrum consists of the numbers~$W(n)$ of $\wlen$-mers occurring 
exactly~$n$ times for all $n=0,1,2,\dotsc$ 
In order to fit the distribution's parameters,
the spectrum $\Bigl(W(n)\colon n=0,1\dotsc\Bigr)$ was considered as 
a set of binned values 
for independently drawn samples
from a continuous DPL distribution:
$W(n)$ was compared to 
the predicted value
$4^{\ell}\Bigl(F(n+\tfrac12)-F(n-\tfrac12)\Bigr)$.
We used custom-made programs to carry out the parameter fitting,
using the Levenberg-Marquardt algorithm~\cite{NR}, a nonlinear
least-squares method, for which the 
starting parameter values were set by likelihood maximization~\cite{ReedJorgensen}.

We defined CpG content of a word~$w$ (Fig.~\ref{fig:CpG}) as the number of 
non-overlapping CG and GC dinucleotides in~$w$. 
The contribution of different annotations (Fig.~\ref{fig:repeats})
were computed by multiplying each~$W(n)$ value in the spectrum 
by the fraction of occurrences within the annotated regions for words appearing~$n$ times
in the entire sequence. 
For the random shuffling of Fig.~\ref{fig:random}, we partitioned the sequence into 
contiguous segments containing exactly 1000 non-ambiguous nucleotides.
Non-ambiguous nucleotides were garbled in each segment by generating a 
uniform random permutation.

Human sequences (original and repeat-masked) and repeat annotations (Figures~\ref{fig:random}, 
\ref{fig:hs12},\ref{fig:CpG} and~\ref{fig:repeats})
were obtained from 
the UCSC genome browser~\cite{GoldenPath} gateway's FTP server (\url{ftp://hgdownload.cse.ucsc.edu/}),
for version hg18 (NCBI Build 36.1). 
(The repeat annotations were generated by the programs
RepeatMasker~\cite{RepeatMasker} and Tandem Repeats Finder~\cite{TRF}.)
Other sequences (Fig.~\ref{fig:prok}) 
were downloaded from 
the NCBI FTP server (\url{ftp://ftp.ncbi.nlm.nih.gov/genomes/}). 
}

%We counted the occurrence of DNA words on one strand only, with the exception of 
%16-mers in the human genome, where both strands were considered. 

%\bibliographystyle{abbrv}
%\bibliography{journals,sequencing,repeats,pseudogenes,comparison,segmentation,genecontent}

\begin{thebibliography}{10}

\bibitem{TRF}
G.~Benson.
\newblock Tandem repeats finder: a program to analyze {DNA} sequences.
\newblock {\em Nucleic Acids Research}, 27(2):573--580, 1999.

\bibitem{genome.bee}
The Honey Bee Genome Sequencing Consortium.
\newblock The genome of a highly social species, the honey bee {A}pis
  mellifera.
\newblock 2006.
\newblock Under revision.

\bibitem{GoldenPath}
A.~S. Hinrichs, D.~Karolchik, R.~Baertsch, G.~P. Barber, G.~Bejerano,
  H.~Clawson, M.~Diekhans, T.~S. Furey, R.~A. Harte, F.~Hsu,
  J.~Hillman-Jackson, R.~M. Kuhn, J.~S. Pedersen, A.~Pohl, B.~J. Raney, K.~R.
  Rosenbloom, A.~Siepel, K.~E. Smith, C.~W. Sugnet, A.~Sultan-Qurraie, D.~J.
  Thomas, H.~Trumbower, R.~J. Weber, M.~Weirauch, A.~S. Zweig, D.~Haussler, and
  W.~J. Kent.
\newblock The {UCSC} genome browser database: update 2006.
\newblock {\em Nucleic Acids Research}, 34:D590--598, 2006.

\bibitem{tinkering}
F.~Jacob.
\newblock Evolution and tinkering.
\newblock {\em Science}, 196(4295):1161--1166, 1977.

\bibitem{Karev.BDIM.2002}
G.~P. Karev, Y.~I. Wolf, A.~Y. Rzhetsky, F.~S. Berezovskaya, and E.~V. Koonin.
\newblock Birth and death of protein domains: a simple model of evolution
  explains power law behavior.
\newblock {\em BMC Evolutionary Biology}, 2:18, 2002.

\bibitem{Karlin.symposium}
S.~Karlin.
\newblock Statistical signals in bioinformatics.
\newblock {\em Proceedings of the National Academy of Sciences of the USA},
  102(38):13355--13362, 2005.

\bibitem{LiKaneko}
W.~Li and K.~Kaneko.
\newblock Long-range correlation and partial $1/f^\alpha$ spectrum in a
  noncoding {DNA} sequence.
\newblock {\em Europhysics Letters}, 17:655--660, 1992.

\bibitem{Luscombe.power}
N.~M. Luscombe, J.~Qian, Z.~Zhang, T.~Johnson, and M.~Gerstein.
\newblock The dominance of the population by a selected few: power-law behavior
  applies to a wide variety of genomic properties.
\newblock {\em Genome Biology}, 3(8):research0040.1Ð0040.7, 2002.

\bibitem{Mantegna}
R.~N. Mantegna, S.~V. Buldyrev, A.~L. Goldberger, S.~Havlin, C.-K. Peng,
  M.~Simons, and H.~E. Stanley.
\newblock Systematic anaysis of coding and noncoding {DNA} sequences using
  methods of statistical lingustics.
\newblock {\em Physical Review E}, 52(3):2939--2950, 1995.

\bibitem{MartindaleKonopka}
C.~Martindale and A.~K. Konopka.
\newblock Oligonucleotide frequencies in {DNA} follow a {Y}ule distribution.
\newblock {\em Computers \&\ Chemistry}, 20(1):35--38, 1996.

\bibitem{universal.expansion.randomization}
P.~W. Messer, M.~L{\"a}ssig, and P.~F. Arndt.
\newblock Universality of long-range correlations in expansion-randomization
  systems.
\newblock {\em Journal of Statistical Mechanics}, 2005.
\newblock P10004.

\bibitem{long.range.DNA}
C.~K. Peng, S.~V. Buldyrev, A.~L. Goldberger, S.~Havlin, F.~Sciortino,
  M.~Simons, and H.~E. Stanley.
\newblock Long-range correlations in nucleotide sequences.
\newblock {\em Nature}, 356:168--170, 1992.

\bibitem{NR}
W.~H. Press, S.~A. Teukolsky, W.~V. Vetterling, and B.~P. Flannery.
\newblock {\em Numerical Recipes in C: The Art of Scientific Computing}.
\newblock Cambridge UniversIty Press, second edition, 1997.

\bibitem{Reed2004}
W.~J. Reed and B.~D. Hughes.
\newblock A model explaining the size distribution of gene families.
\newblock {\em Mathematical Biosciences}, 189:97--102, 2004.

\bibitem{ReedJorgensen}
W.~J. Reed and M.~Jorgensen.
\newblock The double {P}areto-lognormal distribution --- a new parametric model
  for size distributions.
\newblock {\em Communications in Statistics: Theory and Methods},
  33(8):1733--1753, 2004.

\bibitem{ReinertSchbathWaterman}
G.~Reinert, S.~Schbath, and M.~S. Waterman.
\newblock Probabilistic and statistical properties of words: An overview.
\newblock {\em Journal of Computational Biology}, 7(1/2):1--46, 2000.

\bibitem{Pyknon}
I.~Rigoutsos, T.~Huynh, K.~Miranda, A.~Tsirigos, A.~McHardy, and D.~Platt.
\newblock Short blocks from the noncoding parts of the human genome have
  instances within nearly all known genes and relate to biological processes.
\newblock {\em Proceedings of the National Academy of Sciences of the USA},
  103(17):6605--6610, 2006.

\bibitem{multimedia.self-similar}
Z.~Sahinoglu and S.~Tekinay.
\newblock On multimedia networks: self-similar traffic and network performance.
\newblock {\em IEEE Communications Magazine}, 37(1):48--52, 1999.

\bibitem{hs12.finish}
S.~E. Scherer et~al.
\newblock The finished {DNA} sequence of human chromosome 12.
\newblock {\em Nature}, 440(7082):346--351, 2006.

\bibitem{RepeatMasker}
A.~F.~A. Smit, R.~Hubley, and P.~Green.
\newblock Repeatmasker open-3.0, 1996--2004.
\newblock \url{http://www.repeatmasker.org}.

\end{thebibliography}

 \paragraph{Correspondence} Correspondence
should be addressed to M.Cs.~(www: \url{http://www.iro.umontreal.ca/~csuros/}).

\end{document}